\documentclass{revtex4}

\usepackage{graphicx}
\begin{document}
\bibliographystyle{revtex}


\begin{figure}
\leftline{
\includegraphics[width=.20\textwidth,angle=0]{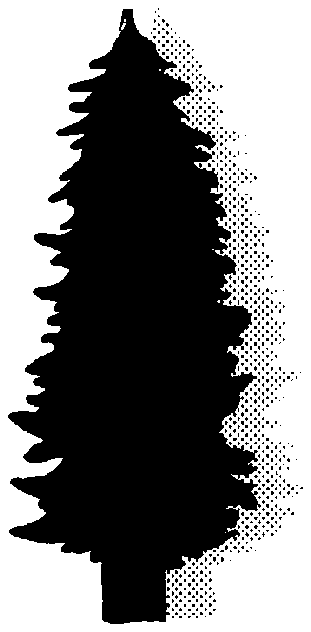}}
\includegraphics[width=.20\textwidth,angle=0]{nothing.ps}
\end{figure}

\vspace*{-10.0cm}
\rightline{SCIPP 01/33}
\rightline{SNOWMASS E3003}
\vspace*{4.8cm}

\title{Research and Development towards a Detector for a High Energy
     Electron-Positron Linear Collider}



\author{Bruce A. Schumm}
\email[]{schumm@scipp.ucsc.edu}
\homepage[]{http://scipp.ucsc.edu/~schumm}
\affiliation{Santa Cruz Institute for Particle Physics and the University of
   California, Santa Cruz}


\date{\today}

\begin{abstract}
This exposition provides a detailed picture of ongoing and planned
activities towards the development of a detector for
a high-energy Linear Collider.
Cases for which research and development activity does not exist, or
needs to be bolstered, are identified for the various subsystems. The
case is made that the full exploitation of the potential of a high-energy
Linear Collider will require the augmentation of existing detector
technology and simulation capability,
and that this program should become a major focus of the
worldwide particle physics community should the construction of a Linear
Collider become likely.
\end{abstract}

\maketitle

\begin{center}
{\small Talk presented at the 2001 Snowmass Workshop on the
Future of Particle Physics \\}
{July 1 -- July 20, 2001}
\end{center}


\vfill
\eject
\setcounter{page}{1}

\section{Introduction}

Over the decade of the 1990's, the five LEP and SLC detectors made
numerous precise measurements of Standard Model couplings and parameters.
Because of this, and the relative cleanliness of $e^+e^-$ beam
collisions, there is a tendency to think that designing and building
a detector for a high-energy (TeV range) $e^+e^-$ Linear Collider (LC)
should be relatively straightforward, requiring an R\&D program of
a scale much smaller than that of its hadron-collider counterparts.
There are a number of reasons to question this point of view.

Although it has been shown that the LC does have unique discovery potential
relative to the LHC,
the physics case for the LC is predicated largely on its ability to
complement discoveries made at hadron colliders with comprehensive and
precise measurements. Over the last few years, it has become
increasingly clear that the full exploitation of the potential of the
LC to do these exacting studies places demands on the design of a LC
detector which can not be met with existing experimental approaches.

For example, the determination of the
light Higgs branching fractions, in particular
into $c {\bar c}$ pairs, benefits from a tracking system with
substantially better vertexing (for flavor identification) and
momentum resolution (for recoil mass-peak cleanliness) than
that of existing detectors. The desire to reconstruct $W$ and $Z$
bosons via their hadronic decays places stringent demands on
the performance of the calorimeter system, suggesting innovative
approaches allowing energy-flow reconstruction, with the requirement
of a large volume of minutely segmented calorimetry, and a correspondingly
large number of channels. No longer a purely $s$-channel machine,
the LC community will also need to think very hard about robust
instrumentation for the far forward region (200 mrad and below).
Finally, the possibility of a fully open trigger, in which all
beam crossings are read out, provides an interesting challenge for
electronics and software efforts, while allowing a number of
interesting physics capabilities. In all of these cases, the full
capabilities of the LC can not be exploited with existing technology.

Underlying all of these potential technological pursuits lies a common
question: that of whether or not the effort and expense engendered by
each individual development project is justified by the
potential physics gain. This set of questions
can only be answered via
a vigorous program of realistic simulation study. It seems imperative
that existing simulation efforts be substantially enhanced, with the
thought in mind that the simulation results should inform and focus
the hardware R\&D program.

\section{The Nominal Linear Collider Detectors}

There are currently four cylindrical-geometry detector scenarios
under consideration for the TeV-scale Linear Collider (a fifth,
so-called `Precise' detector under consideration for a moderate-energy
IP will not be discussed here)\cite{tdrs}. All four detector schemes incorporate
a precise pixel-based stand-alone inner tracker to accurately
reconstruct charged track impact parameters. 
Precise momentum and angular
resolution
is provided by a large-volume central tracker, implemented either as
a drift chamber (`DC'; the Asian, or `JLC' design), a time-projection
chamber (`TPC'; the TESLA and North American `L' designs), or a
larger area solid-state tracker (`SST'; North American `SD' design).
The radius of the
first sensitive layer varies from 1.2 cm (L, SD) to 2.4 cm (JLC), while
the full tracking systems are immersed in axial magnetic fields of between
2 (JLC low-field option) and 5 (SD) Tesla.

Proposed electromagnetic calorimeter solutions include either a highly-pixellated
silicon-tungsten sandwich design (TESLA and SD), or a lead-scintillator
sandwich design (JLC, L). The TESLA collaboration is also considering a
Shashlik option. Proposals for hadronic calorimetry include lead-scintillator
(JLC, L, SD), and iron-scintillator (TESLA). TESLA has also recently introduced
the notion of `digital' hadronic calorimetry, for which showers initiated
in the iron absorber material would be read out in terms of a simple yes/no
response from highly segmented ($\sim$ 1 cm$^2$) gas chambers.

Forward tracking systems with coverage down to approximately 100 mrad
are envisioned for the TESLA, L, and SD detectors, with silicon-strip
disks being the most likely technology. Forward tagging calorimetry
down to 30 mrad or less is envisioned in all cases, with possible
implementations including silicon-diamond and an innovative solid-state
`3-D Pixel' design.

\begin{figure}
\includegraphics[width=.80\textwidth,angle=0]{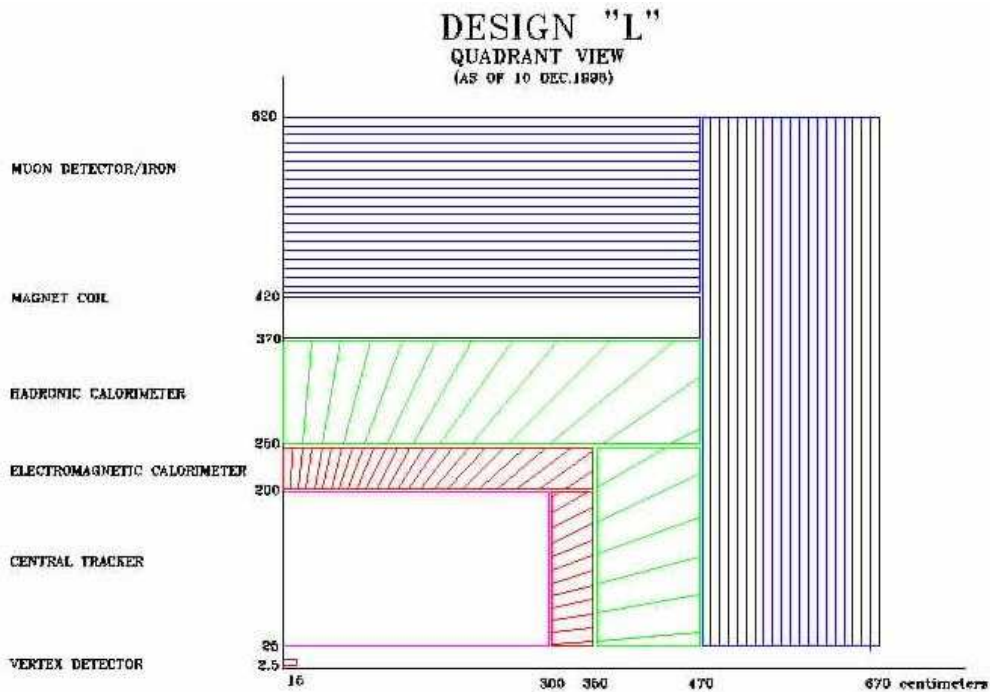}
\caption{Schematic of the North American Large (`L') Detector Design}
\label{detfig}
\end{figure}

Figure~\ref{detfig} shows a schematic of the North American L detector, typical of
the proposed designs.

\section{Tracking}

The motivation to improve the precision of charged-particle tracking
for Linear Collider detectors has been driven by several factors.
To begin with, the need to do flavor -- and particularly charm --
separation has driven a push towards more precise impact parameter
reconstruction and vertexing. For example, the prospect of a light
Higgs motivates an interest in the absolute $H \rightarrow c {\bar c}$
branching fraction, while the effective luminosity of the study of
symmetry breaking via strong coupling between longitudinal $W$ boson states
can be enhanced by a factor of two or more via the use of a
clean tag of charm in the $W$ decay jet. In both of these cases,
the impact parameter performance of existing vertex detectors would
not be adequate to extract the full physics capabilities of the Linear
Collider.

In addition, in order to identify the Higgs signal, and
constrain the Higgs width,
the reconstruction of the missing-mass spectrum from the critical channel
$e^+e^- \rightarrow Z H$; $Z \rightarrow \mu^+\mu^- (e^+e^-)$
should not degrade the $0.2\%$ width expected from the beam energy spread.
In addition, the dilepton mass resolution should be small compared to the
natural width of the $Z$, even at the large $Z$ kinetic energies that
will be associated with this `Higgstrahlung' process at the highest
achievable machine energies.
This requires that the overall momentum of individual tracks be reconstructed
with an accuracy approaching $\delta p / p^2 \simeq 1 \times 10^{-5}$,
representing an improvement of over an order of magnitude relative
to existing $e^+e^-$ detectors. While a large portion of the improvement
is expected to come from increasing the strength of the magnetic field,
achieving this degree of accuracy will also require a large radial lever
arm, good single-hit resolution, and careful attention to alignment and
calibration systematics.

\subsection{Precision Vertexing}

\begin{figure}
\includegraphics[width=.80\textwidth]{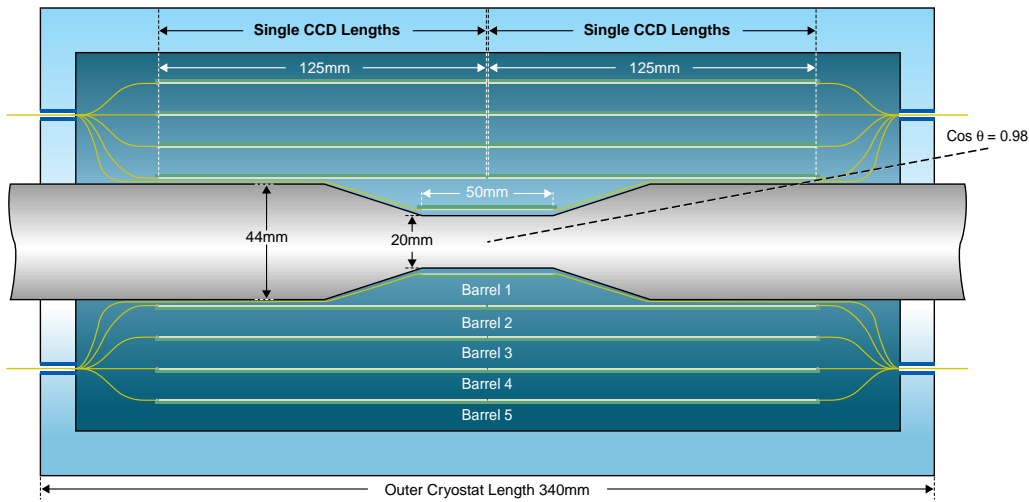}
\caption{Design schematic for the common TESLA and North American
Linear Collider Vertex Detector.}
\label{vxd}
\end{figure}

\begin{figure}
\includegraphics[width=.60\textwidth]{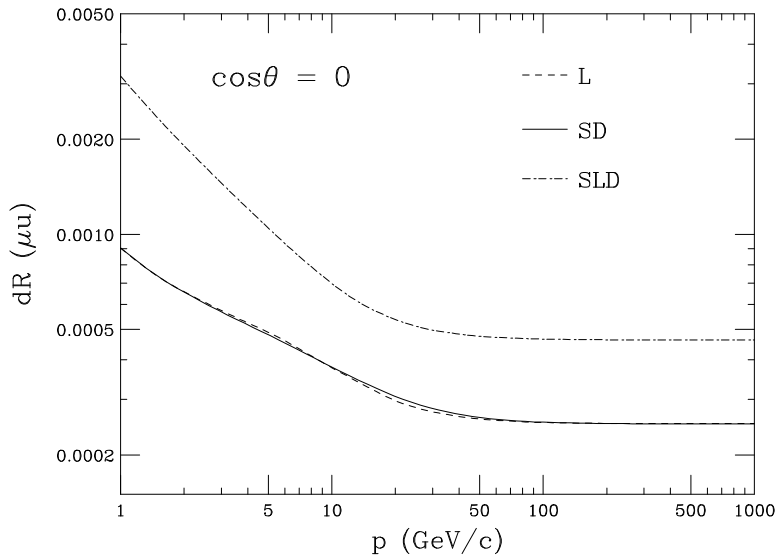}
\caption{Impact parameter performance of the Linear Collider Vertex
Detector. To the extent that the momentum resolution is similar, this
performance is independent of the choice of central tracker (TPC for
`L' versus solid-state for `SD'. The performance of the former SLD
system is shown for comparison.}

\label{vxres}
\end{figure}

The SLD VXD3~\cite{VXD3} vertex detector has set the current standard for
precision tracking at $e^+e^-$ colliders, and the common VXD design
for the TESLA, L, and SD detector scenarios are essentially an outgrowth
of that successful design. The Linear Collider VXD design, shown in
Figure~\ref{vxd}, is a pixel based stand-alone tracking system
with five layer coverage to $|\cos\theta| = 0.87$ and single-layer
coverage out to $|\cos\theta| = 0.98$. It is expected to achieve
an $r-\phi$ impact parameter resolution at high momentum of better
than 3 $\mu$m, with a multiple scattering term of less than 10 $\mu$m
at $p_{\perp}/\sqrt{\sin\theta} = 1$ GeV/c. Both of these are substantially
better than those achieved with the SLD VXD3, as can be seen in Figure~\ref{vxres}.

The baseline VXD design is instrumented with
$20 \times 20$ $\mu$m$^2$ CCD pixels,
demanding approximately $7 \times 10^8$ pixels, only about a factor of
two greater than that of the existing VXD3 detector of the SLD.
However, due to the need to maintain signal efficiency over the
large number of transfers from the struck pixel to the readout
pad, CCD's are relatively sensitive to radiation damage. For expected
neutron fluences of order $10^9$ cm$^{-1}$, the charge-transfer efficiency
becomes significantly compromised by the development of traps. It
has been shown~\cite{ccdrad} that filling these traps, for example with
carriers released in response to a flash of light, largely restores the
charge-transfer efficiency for a period approaching one second. While
an encouraging proof of principle, this notion needs to be engineered into
a working solution.

The greater luminosity of the TeV-scale Linear Collider will require
an increase in readout rate from 5 to 50 MHz, in order to maintain the
CCD occupancy low enough that it will not compromise the physics at
a 120 Hz JLC/NLC. For TESLA, however, integrating over
the several thousand crossings of the 800 $\mu$s-long
TESLA bunch train will lead to prohibitive occupancy in the CCD's,
even with 50 MHz readout.
Thus, the design is underway for a CCD sensor which will allow parallel
readout of the individual detector columns (`column-parallel' readout),
allowing the detector to be read out many times during the 800 $\mu$s spill.

The VXD baseline design calls for a ladder thickness of 0.12\% $X_0$ or
thinner --
a substantial reduction from the 0.4\% $X_0$ achieved for the SLD VXD3.
Achieving this goal will require the ladders to be supported from the
ends only; substantial work is getting underway to explore the
design and stability of appropriate support systems. In addition,
it may be possible to improve upon the 3 $\mu$m single-hit resolution
that has already been achieved with 20 $\mu$m CCD pixels. Again, the
motivation for these various R\&D paths needs to be provided via
the simulation of physics channels which may depend upon them.
Finally, while a substantial and formal R\&D task force already exists
(the UK Linear Collider Flavour Identification Collaboration), the
amount of work to be done towards a final design is substantial, and
is by no means saturated by this group.

An interesting alternative to CCD pixels, particularly if the radiation
damage or readout issues turn out to be insurmountable, is that of
monolithic CMOS pixel detectors\cite{CMOS}.
Rather than having the local electronic
circuitry bump-bonded to the pixel sensors, as for conventional
fast pixel detectors, the first-stage readout
is deposited directly onto the sensor wafer, allowing for
a much thinner structure than that of standard active pixel arrays.
In addition, the monolithic approach seems fairly promising in terms of
achieving the small ($20 \time 20$ $\mu$m$^2$) pixel dimension required
to avoid problems with cluster merging at the TeV-scale Linear Collider.
The active pixel group at IReS Strasbourg is developing a monolithic
sensor for which a local array of three
transistors will amplify and drive signals from $20\times20 \mu$m$^2$ pixels,
which can be read out individually via shift-register controlled column and
row selectors. Test beam results with early prototypes suggest high
hit efficiency ($\sim 99\%$) and good resolution ($\sim 1 \mu$m).
Issues being addressed with current development work include the development
of large-scale (detector-sized) arrays, system and integration issues, and
radiation hardness.

\subsection{Central Tracking}

Proposals for LC central tracking systems fall into two main categories: gaseous
and solid-state. Proposals for gaseous tracking systems include a large-volume
TPC (TESLA and North-American L), and a mini-jet-cell drift chamber with a longitudinal
extent of greater than 4 meters (JLC). The North-American SD design calls
for a large-area solid state tracker, implemented either with silicon strip
or silicon drift technology. A substantial R\&D effort is called for in order
to develop and optimize each of these systems for the LC detector.

The Japanese drift chamber development program has been driven primarily
by the chamber length, as well as the large magnetic field
needed to achieve a resolution of $\delta(1/p_{\perp})$ of $10^{-4}$. They
have addressed issues associated with electrostatic and gravitational wire
sag, stereo wire geometry, wire tension loss, and the large Lorentz angle.
Understanding the sensitivity to LC backgrounds, including neutrons, is a
critical issue that still needs to be addressed.

A sizable collaboration, mostly European but also
including several North American institutions,
is exploring the use of a large-volume TPC at the LC. The TPC enjoys several
nominal advantages over a conventional drift chamber, including true three-dimensional
track measurements to help disentangle dense jets, as well as a
high degree of pixellation to reduce sensitivity to LC backgrounds. The
goal of $\delta(1/p_{\perp}) \simeq 5 \times 10^{-5}$
places substantial demands
on the spatial resolution of the endplate readout, as well
as its ion feedback suppression capability. In addition, alignment tolerances
must be maintained to several microns -- substantially better than that of
existing TPC's -- requiring the exploration of calibration and environmental
control techniques.

A large effort is currently focussed on the development of micro-pattern
gas detector technology (GEM and MicroMEGAS),
which would provide an alternative to the standard MWPC readout systems.
Studies suggest that optimized readout pad structure will admit appreciable
improvements in both the single-hit and two-track resolution relative to
those of the MWPC readout. In addition, the natural ion feedback suppression
of micro-pattern detectors may allow the elimination of the gating grid, and
the corresponding need to run the TPC in a triggered mode.

A number of other R\&D issues are being explored for the TPC option. The gas
mixture is being studied in order to optimize tradeoffs between resolution versus
ion clearing, and quenching versus neutron background susceptibility. The emphasis
on LC physics in the forward region is somewhat new for $e^+e^-$ detectors, and
thought must be given to minimizing the thickness in the endplate region.
Finally, the high degree of spatial and temporal pixellation desired for the
TPC readout provides a challenge to the development of fast multichannel electronics.

Both silicon-strip and silicon-drift are now mature tracking technologies, an
example of a system composed of the latter being the STAR collaboration's
silicon vertex detector\cite{STAR},
for use in studying charged tracks from heavy ion collisions at RHIC.
The baseline design of the North American SD detector calls for five
cylindrical layers of solid-state tracking with radii between 20 and
125 cm, and a polar angle coverage of $|\cos\theta| \le$ 0.8.
Either option (strip or drift) seems applicable to such a design,
although again a substantial R\&D program
would be required in order to optimize the design of either.

A nominal advantage of solid state over gaseous tracking arises in the
context of background immunity, due to the fact that the energy deposition
of low-energy gamma conversions is very localized in the dense silicon.
In the case of silicon drift, the detector is intrinsically
three-dimensional (due to the drift of the liberated conduction electrons
in the $z$ direction), leading to further reduction of background
sensitivity via the correspondingly fine pixellation. 

\begin{figure}
\bigskip
\centerline{
\includegraphics[width=.45\textwidth,angle=90]{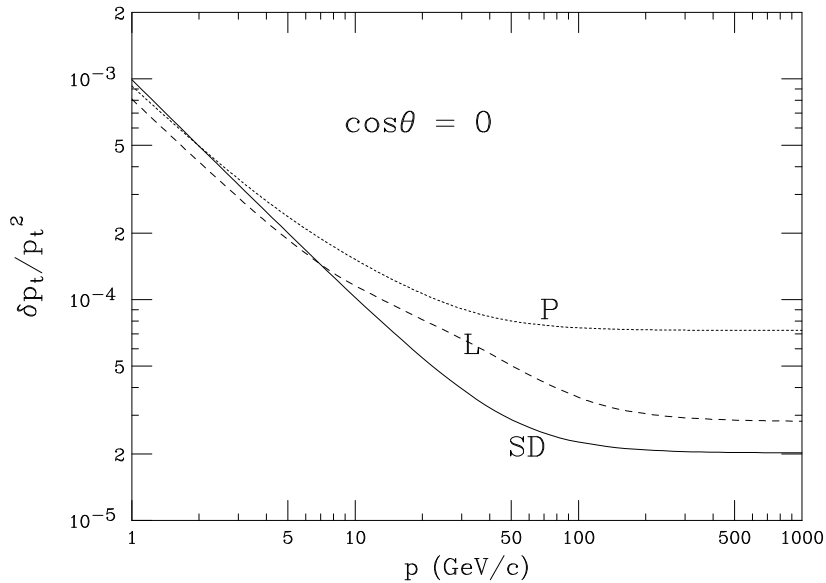}}
\caption{Momentum resolution versus momentum at $\cos\theta = 0$
for various North American detector
designs (the performance of the TESLA and Asian designs are similar).
P = Precise; L = Large; SD = silicon.}
\label{ptres}
\end{figure}

Two mutually exclusive approaches have been
proposed for the design of the SD silicon strip detector. 
With the development
of short shaping-time readout, it may be possible to time-stamp silicon strip
hits with a resolution of as little as 5 nsec, 
leading to
a corresponding increase in the ability to select against LC backgrounds.
On the other hand, long shaping-time readout offers the possibility
of long, thin detector ladders which can be read out without introducing
electronics into the detector fiducial volume, leading to superior
$p_t$ resolution even at low momentum. In addition, the large (125 cm)
radial lever arm and high (5T) field of the SD detector yield, assuming an
$r$-$\phi$ resolution of 7 $\mu$m, a resolution in $\delta p_t/p_t^2$
approaching $2 \times 10^{-5}$. A comparison between the North American L,
SD, and P detectors (the latter not discussed here)
of the momentum resolution vs. momentum, at $\cos\theta = 0$, is shown in
Figure~\ref{ptres}.

A number of other hardware issues more or less common to both solid state
detector options need to be addressed. If the low LC collision duty cycle
can be taken advantage of by the readout electronics, it should be possible
to manage local power consumption with passive cooling. 
The development of a frequency-scanned interferometric
alignment system for the ATLAS inner detector\cite{ATLAS} suggests that
the maintenance of such tolerances are within reach of current technology.
The effects of the large magnetic field on the drift of electrons and
holes must be understood and controlled. Finally, comprehensive
radiation damage studies need to be conducted, in order to ensure that the
proposed solid state detectors (particularly for the case of silicon drift)
can withstand LC backgrounds.

\subsection{Forward Tracking}

Much of the interesting physics at a high energy LC is not produced
via the $s$ channel, and as such tends to be boosted into the far
forward or backward regions of the detector. Thus, substantially
more emphasis needs to be given to these regions of the detector than
was for the case of the $Z^0$-pole $e^+e^-$ detectors. The forward tracking
system needs to maintain good momentum resolution for the accurate measurement
of the slepton spectrum endpoint, and fractional resolution of the
dip angle approaching one part in $10^4$, in order to appropriately
constrain the differential luminosity spectrum via the asymmetry in
radiative bhabha events.

The North American designs have preserved momentum resolution of better than
$\delta p_t / p_t^2 = 7 \times 10^{-4}$ out to $\cos \theta = 0.99$ with
a system of silicon-strip $r$-$z$ disks. The TESLA design improves this
to $\sim 2 \times 10^{-4}$ with an additional
fine-grained forward tracking chamber
(FCH) just beyond the TPC endplate. Little effort has been put into the
design of forward tracking systems other than the optimization of the
overall detector geometry that can be done with closed-form track parameter
error calculations. The detailed design and integration of a precision
forward tracking system that can withstand the low-angle LC backgrounds
remains a wide-open and critical R\&D issue in the
development of the LC detector.

\subsection{Tracking Simulation Issues}

Substantial thought has been put into the proposal of approaches that
minimize tracking resolution parameters, with the goal of producing a
LC detector design that can exploit the intrinsic precision available
from high energy $e^+e^-$ collisions. While some simulation work has been
done to explore the extent to which tracker parameters, such as
point resolution, radial lever arm, and material burden,
need to be pushed in order to exploit
LC physics, much work still remains to be done. Eventually, choices between
detector technology, or of whether or not to pay for an incremental
improvement in a given technology, will need to be based on a concrete
assessment of the physics advantage to be gained.

Studies underway to understand the impact of vertexing parameters such
as point resolution, layer thickness, and inner radius on the measurement
of the Higgs branching fraction and strong $WW$ scattering parameters need
to be brought to fruition. Detailed studies of the dependence of the
selectron mass measurements on the forward tracking resolution need to
be undertaken. Physics channels need to be simulated which will establish
targets for the polar angle resolution necessary to perform the differential
luminosity spectrum. Targets for transverse momentum resolution,
both a low and high momentum, need to be established in order
to inform R\&D programs directed at lessening the material burden and
improving the accuracy of the sagitta measurement. 

Thought should also be
given to the advantage of having dE/dX information, which would be of
higher quality for a gaseous tracker. While $\sim 10$ layers of silicon
(vertexer plus SD central tracker) do provide a reasonably accurate
energy loss measurement, the density effect limits the differentiation
between particle species in the relativistic rise region. Thus, for
relativistic electrons which do not reach the calorimeter, dE/dX information
from a gaseous tracker may well be the only to provide separation from pions.
The extent to which the consideration of dE/dX resolution is relevant to the design
of the LC detector is an issue
that needs to be addressed by the simulation effort.

While the aforementioned issues are probably best addressed with a fast
simulation, a number of more detailed questions can only be addressed
with a full-scale detector simulation and realistic reconstruction.
Concerns that the limited number of widely spaced hits from a solid
state tracker might compromise the ability to distinguish closely
spaced tracks in collimated jets, or to detect the decay kinks of long-lived
exotic states, need to be addressed with this more realistic simulation.
Another interesting question that can be addressed with the full simulation
is the need for a central tracker with intrinsically three-dimensional
tracking, such as that provided by a TPC or silicon drift detectors, in view
of the ultra-fine vertex detector pixellation. No study has been done
as of yet regarding the relative track-separation resolutions of the
various proposed solutions. The value of temporal pixellation, in regards to
background suppression, needs to be explored, as does the value of
an intermediate silicon tracker in the gaseous detector options.
Finally, the entire issue of the interplay between the tracker and calorimeter,
in terms of backsplash into the outermost layers, as well as optimizing
the energy flow measurement (to be discussed below), remains to be
explored with the full simulation.

\section{Calorimetry}

Precise, hermetic, full-coverage calorimetry is thought to be
a must for a LC detector. The ability to make precise
jet energy measurements for hadronic final-state reconstruction,
to have a discriminating tool for electron-pion separation,
and to perform measurements of neutral particle trajectories
with a precision approaching that of a crude tracking system,
seem to be desirable goals for the LC detector calorimeter design.
Accumulated wisdom suggests that the way to accomplish this
is with `energy-flow' calorimetry.

\begin{table}
\caption{Measurement of the components of high energy jets}
\label{eftab}
\smallskip
\begin{tabular}{|c|c|}
   \hline
  {\bf Component}   &    {\bf Measurement}    \\  \hline  \hline
  Neutral pions   &   $\sim 15\%/\sqrt{E}$ in EMCAL  \\
  Photon          &   $\sim 15\%/\sqrt{E}$ in EMCAL  \\
  Charged hadrons &   $\sim 0.1\%$ in tracker    \\
  Charged leptons &   $\sim 0.1\%$ in tracker     \\
  Stable neutral hadrons &   $\sim 40\%/\sqrt{E}$ in HADCAL \\
  Neutral leptons &   No measurement                     \\  \hline
\end{tabular}
\end{table}

The idea behind calorimetric energy-flow measurement is encapsulated
in Table~\ref{eftab}. All components of hadronic jets are
well-measured except a relatively small fraction of stable
neutral hadrons (mostly $K_L$ and neutrons). If the calorimetric
energy deposits from charged hadrons can be removed from the jet,
and replaced with the amply precise tracking measurement, overall
jet energy resolution can approach that of the main remaining
calorimetric component, which is electromagnetic and thus
relatively well-measured. Clearly, achieving this goal
places stringent requirements on the track-cluster association
capabilities of the combined calorimeter and tracking system.

Toy MC studies\cite{rayfrey} performed with a JETSET sample of
$e^+e^- \rightarrow q{\bar q}$ events indicate that, in the ideal
situation that every charged hadron contributes to the jet measurement
only through its tracking information, a jet energy resolution of
$18\%/\sqrt{E}$ is achieved. On the other hand, with perfect
$e/h$ compensation, and using the calorimeter only, the resulting
jet energy resolution is $64\%/\sqrt{E}$.

\begin{figure}
\includegraphics[width=.40\textwidth]{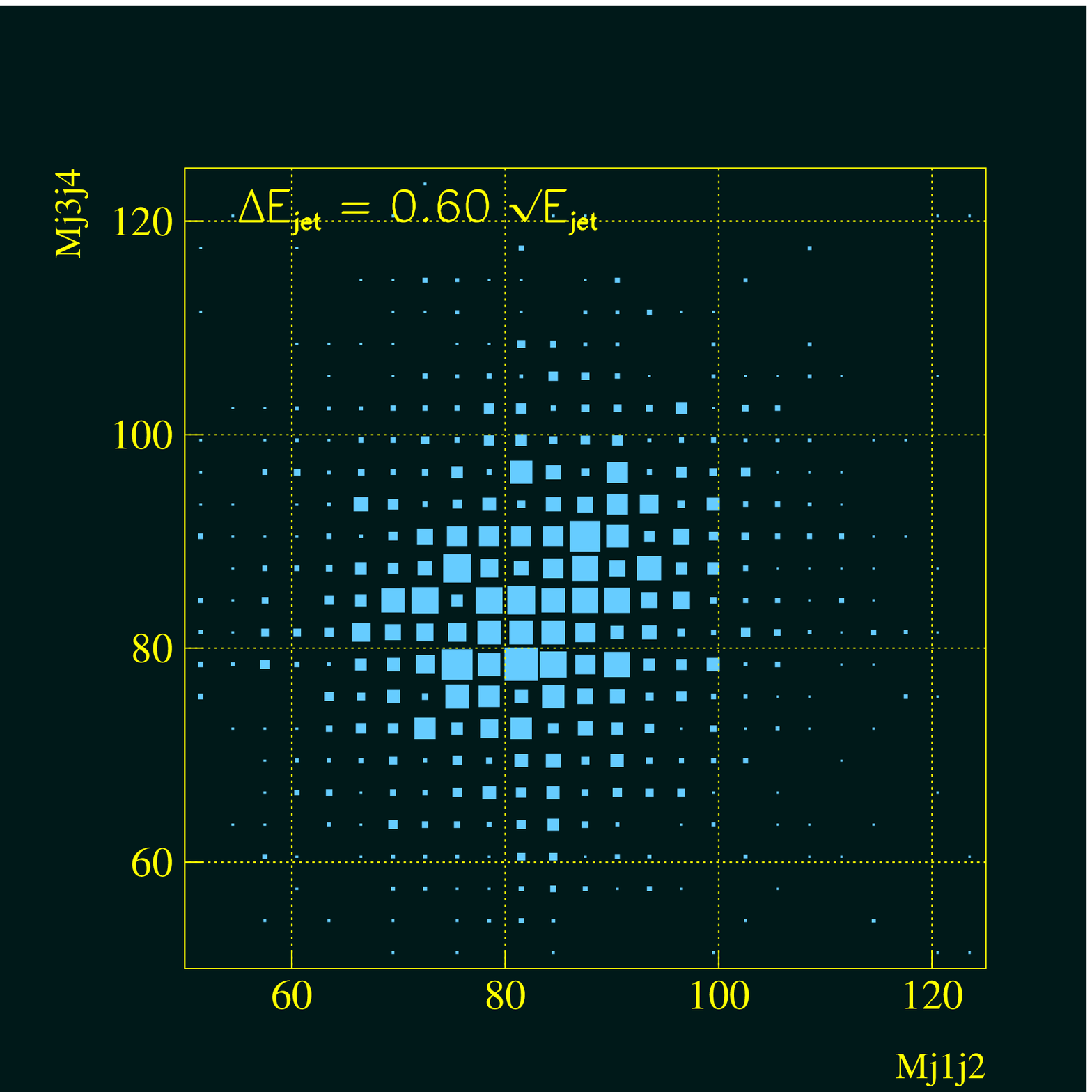}
\includegraphics[width=.40\textwidth]{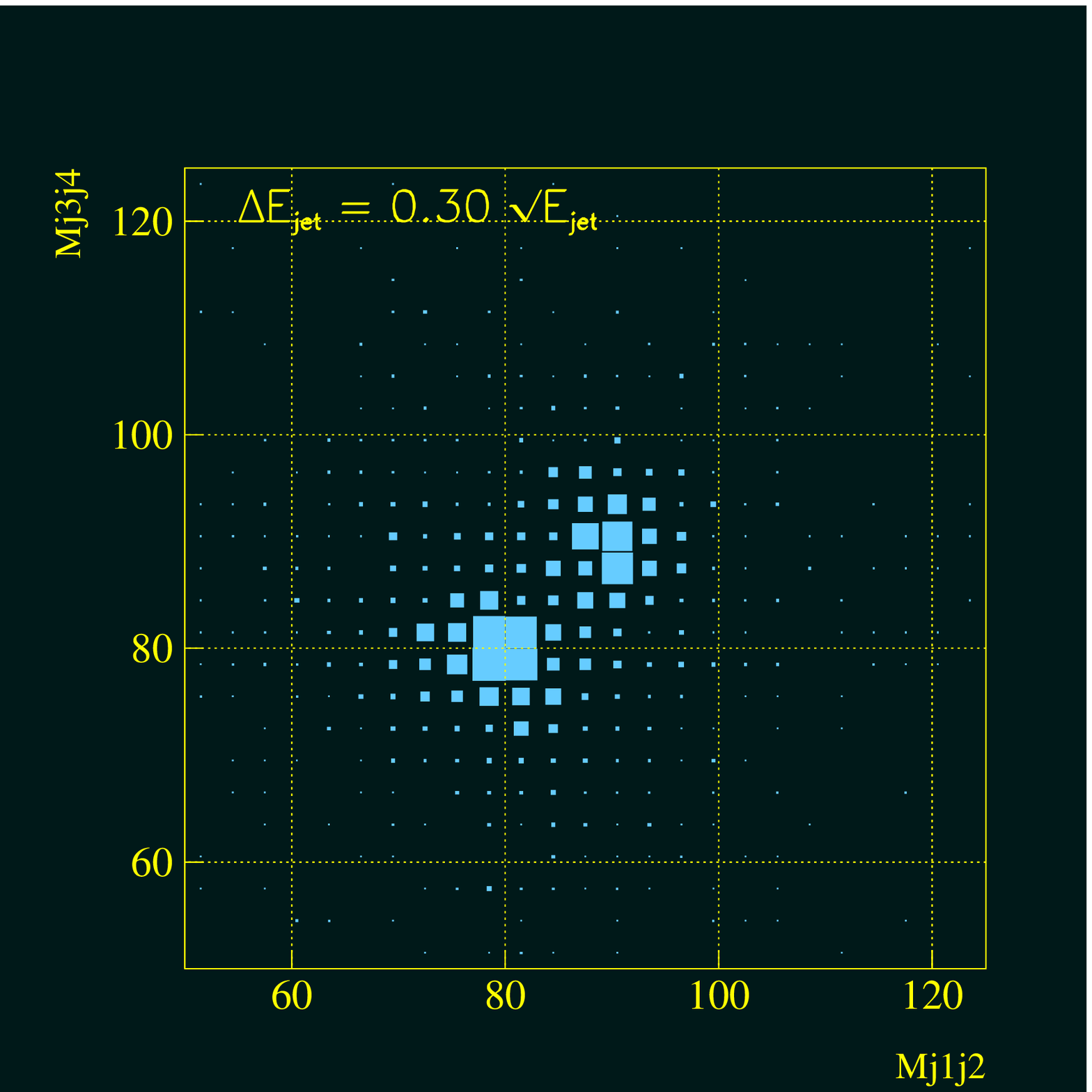}
\caption{Reconstruction of the jets in the processes
$e^+e^- \rightarrow Z^0Z^0$ and
$e^+e^- \rightarrow W^+W^-$ at $E_{cms} = 500$ GeV
for jet energy resolutions of $60\%/\sqrt{E}$ (left) and
$30\%/\sqrt{E}$ (right).}
\label{zwsep}
\end{figure}

Although this ideal jet energy resolution will be difficult to achieve
in practice, even a resolution of $30\%/\sqrt{E}$ will provide a
significant advantage for the isolation of physics signals. An example
of this --
the separation of $e^+e^- \rightarrow ZZ\nu{\bar \nu}$ from
$e^+e^- \rightarrow WW\nu{\bar \nu}$ -- is shown in Figure~\ref{zwsep}.
For a resolution of $60\%/\sqrt{E}$, the dijet mass signal region
mixes the two channels together, while for $30\%/\sqrt{E}$, the
separation of these two channels is distinct.

Generic requirements for energy-flow calorimetry include a large
product of magnetic field and inner radius squared ($BR^2$) to provide
a large separation between charged tracks. In order to keep the
calorimeter showers well contained and separated from neighboring showers,
it is necessary to achieve a minimal Moliere radius, and to have the
fine readout segmentation needed to take advantage of the resulting small
shower spread. Frequent longitudinal sampling will allow the
sharpest discrimination between electromagnetic and hadronic depositions,
as well as the most precise reconstruction of the trajectory of
incident charged particles (for the purpose of track/cluster matching).
Finally, the tracker needs to be well integrated, so that
material in the outer reaches of the tracker does not degrade the
cluster-matching capabilities. The choice of technology to achieve
these goals is currently focussing on a silicon/tungsten sandwich,
with a readout pixellation of order 1 cm$^2$ to complement the Moliere
radius of tungsten.

The greatest progress towards establishing the feasibility of
energy-flow calorimetry in high-energy $e^+e^-$ collisions
has been made by the TESLA collaboration. Among other things,
this work has led to
a novel approach to hadronic calorimetry. With a resolution
of typically $40\%/\sqrt{E}$, the hadronic resolution
tends to dominate the accuracy of the
jet energy resolution. TESLA has proposed very fine
pixellation (again 1 cm$^2$), for an iron/gas sandwich 
hadronic calorimeter that is
read out digitally (yes/no for each cell), 
as an alternative to their previous $5\times 5$ cm$^2$
iron/scintillator baseline, which would be read out in an analog mode.
The idea behind this so-called `digital' hadronic calorimeter is to
continue from the EM calorimeter, to an extent that is economically
feasible, the ability to distinguish electromagnetic sub-clusters in
the hadronic shower. Incidentally, this technology would also allow
muon identification below 5 GeV/c, for which muons do not have enough range
to traverse the calorimeter and enter the muon chambers.

Development is underway for
the more conventional HCAL approach, for which the calorimeter
is instrumented with $5 \times 5$ cm$^2$ scintillator tiles read out via
wavelength shifting fibers. The fibers are bundled to form calorimeter cells,
and read out with photodiodes. The group in currently in the process of
doing lab tests on several industrial alternatives for the various hardware
components of the calorimeter, including the scintillator tiles, wavelength-shifting
fibers, optical wrappings, and optical readout fibers. They are also
exploring the optimal way to couple the scintillation light into the
wavelength-shifting fibers, and transporting the light to and coupling the light
into the photodiodes. The photodiode is also under development, with a
32-channel Avalanche Photodiode pixel array developed at DESY and MPI-Munich
currently under study.
It is expected that this process will continue through
summer 2002, and will result in the construction of a 27 layer prototype
`Minical' for test beam studies.

A decision between these two HCAL alternatives is expected soon -- perhaps
as early as November, 2001 -- and preliminary simulation studies have
been favorable to the `digital calorimeter' approach. Simulations
of isolated pions entering the digital calorimeter have indicated
hadronic energy resolution of $29\%/\sqrt{E}$, maintaining over a wide
range of incident energies.

TESLA has conducted a number of interesting simulation studies on the
trade-offs between cost and performance in the EM calorimeter. The
EM design proposal call for 40 silicon-tungsten layers in about 25 $X_0$.
The group has found very little degradation associated with a dead-wafer
fraction as high as 5\%; the ability to operate with this many bad silicon
wafers would relax production tolerances in a way that may save as much as
a factor of two in production costs. Reducing the number of layers from 40
to 20 would yield the expected degradation of $\sim \sqrt{2}$ (from
10\%/$\sqrt{E}$ to 14\%/$\sqrt{E}$) for individual electromagnetic particles.
However, much of the interesting physics that would be done with calorimetric
information would rely on jet rather than individual particle reconstruction,
for which the fractional degradation would be much less.

In the context of jet reconstruction, simulations of the processes
$e^+e^- \rightarrow Z^0 \rightarrow q{\bar q}$ at $E_{cm} = 91$~GeV,
and $e^+e^- \rightarrow \gamma Z^0 \rightarrow \gamma q {\bar q}$
at $E_{cm} = 500$ GeV, have been conducted. A full-scale simulation
of these processes is very involved, requiring the development and
optimization of realistic track/cluster matching and energy flow
algorithms. In this case, a less sophisticated approach was used,
in which individual particles were reconstructed with ideal
resolutions, but a parameterization of a realistic subtraction/substitution
package was used to replace hadronic clusters with their tracking
information\cite{vasily}.

For a hadronic calorimeter resolution of
$30\%/\sqrt{E}$, these simulations indicated an overall jet-energy
resolution of $26\%/\sqrt{E}$, a value thought to be quite interesting
given the result of Figure~\ref{zwsep}. While these simulations may be
somewhat ideal, it also may be that the track/cluster matching algorithm
is not completely optimal. Thus, it's difficult to say whether these
results are optimistic or pessimistic. A similar study for the
iron/scintillator solution found a jet-energy resolution of
$46\%/\sqrt{E}$, although with the use of a substantially less developed
track/cluster matching algorithm. Nonetheless, the simulation studies
do suggest the possibility of substantially improved performance with
the digital calorimeter approach. Much work -- both in shoring up the
simulation studies as well as presenting a design for a gaseous cell
that can be read out with a pixellation of 1 cm$^2$ -- must be
completed before the choice of this approach can be confidently made.

One important issue that has been raised by the TESLA group's simulation
work is a potential problem with background in the track/cluster association
in the forward sections of the electromagnetic calorimeter. The source
of spurious clusters seems to arise from the albedo associated with
charged hadron interacting in the mask, which lies just inside the
endcap calorimeter, and is used to absorbs backsplash from the
exiting disrupted beam as it re-enters the machine aperture. More
work needs to go into these simulation studies in order to understand how
to keep these backgrounds from compromising the energy-flow capabilities
of the forward calorimetry.

An alternative being considered by the TESLA collaboration, which would
be particularly attractive if the advantages provided by energy-flow
calorimetry are not thought to be central to the LC physics case,
is a Shashlik electromagnetic calorimeter. The Shashlik design incorporates
a lead/scintillator sandwich, with the scintillator read out by a
wavelength-shifting fiber. Transverse pixellation of as little as
3 cm$^2$ can be achieved, while a degree of longitudinal segmentation can
be accomplished via the use of scintillator materials with different decay
times. The Shashlik option offers the potential for a favorable ratio
of performance to cost, although its ability to provide clean track/cluster
association and trajectory information is limited relative to that of
the proposed silicon/tungsten design.

The Asian study group is proposing a conventional lead/scintillator
sandwich whose thicknesses in the hadronic portion are optimized
for e/$\pi$ compensation. The scintillating tiles will be read out
via wavelength-shifting fibers.

The electromagnetic calorimeter is divided
into three sections of 4 $X_0$ (preshower), 9 $X_0$, and
14 $X_0$, respectively, and with a tile dimension of
$6 \times 6$ cm$^2$.
The electromagnetic calorimeter design also incorporates a
1 cm scintillator strip array a shower-max, with 1 cm$^2$ silicon
pads being maintained as an option should studies show it to be
warranted. The hadronic calorimeter consists of four sections of
about 1.5 $\lambda_0$ each, for a total of 6.5 $\lambda_0$. The
transverse segmentation in the hadronic calorimeter is $18 \times 18$
cm$^2$.

The Asian group would like to achieve a two-jet resolution better than
the Z-boson and W-boson widths, to avoid having the reconstruction
of these states limited by detector resolution. A fast MC simulation,
including the use of track-cluster association, suggests a resolution
on W-boson dijets of 2.9 GeV, somewhat worse than the 2.1 GeV
natural width of the W. This simulation, however, did not include the
potentially advantageous use of the shower-max information.

The Asian group has conducted beam tests of their tile/fiber prototype
both at KEK and FNAL. For electrons, they have found
$\sigma_E/E = (24 \oplus 0.8)\%$, while for pions they measure
$\sigma_E/E = (47 \oplus 0.9)\%$. They observed better than 1\% linearity
for the entire explored range between 2 and 150 GeV. Some signal loss was
experienced for electrons entering near tower boundaries, which will
be an avenue of further development work.

The Asian group's continued development work is focussing on a number of
areas. The group is pursuing affordable means for producing reliable
scintillator strips for the shower-max detector. They are also looking into
a number of potential ideas for reading out the strips, including
high-performance multi-channel photon detectors, or even avalanche photodiodes
to read the strips out directly, without the use of wavelength-shifting
light guides. They are also beginning an effort to develop a
full simulation which would allow them to realistically address the study
of jet-energy resolution, given the performance on individual particles
observed with the test beam.

The worldwide LC calorimetry effort could benefit greatly from additional
simulation and prototyping effort. 
While studies such as that of 
Figure~\ref{zwsep} suggest that energy-flow calorimetry can be a very
powerful tool indeed, they fall short of informing us of the overall
programatic advantage of implementing energy-flow capabilities in the LC
detector. It would be beneficial to have a number of realistic
studies of the physics
reach in various essential channels as a function of calorimeter technology.
In addition, should energy-flow calorimetry be thought desirable, full-scale
simulation of the entire measurement system, including the interplay between the
tracking system (reconstruction parameters and material burden) and
calorimetry, will be necessary in order to build confidence in specific designs.
Finally, prototyping of the more promising avenues should begin,
so that well-founded estimates of cost and performance parameters can
be developed.

\section{Triggering and Electronics}

More than its energy reach, the complementarity of the LC program to that
of the LHC makes the LC program interesting and worthwhile. One important
aspect of this complementarity is the openess of LC detector trigger.
A consideration of the event size and data rate suggest that computing
and I/O technology may enable a fully open trigger, for which all relevant
detector information is written out for every beam crossing. In some
scenarios, such as certain gauge-mediated SUSY and universal extra dimension
models which produce signatures containing soft charged tracks and
undetectable long-lived heavy particles, this capability could prove
essential.

In addition to pushing the frontiers of computing technology, the
development of a fully open trigger would require substantial electronics R\&D,
and perhaps even detector physics R\&D, in order to achieve the
necessary degree of zero-suppression. Other aspects of proposed LC
detector options, such as a finely pixellated TPC endplate,
optimized silicon detector readout, and the readout for a highly pixellated
energy-flow calorimeter, will also require substantial electronics
development work.

\section{Other Systems}

There are a number of other systems, potentially of critical importance to the LC
detector and its physics program, whose development will also require
substantial R\&D and design work.

Far-forward calorimetry, from the limit of the endcap coverage down to
within 20 mrad or so of the beam axis, is important for tagging and
vetoing forward-scattered beam particles, in order to remove backgrounds to
selectron studies and signals with large missing energy, and to detect
radiative electrons in studies of two-photon physics. In light of the
intense backgrounds associated with the proximity of the beam, these
detectors will have to be very robust against radiation damage, and
have a high degree of spatial and temporal segmentation. Conventional
silicon/tungsten calorimeters may suffice for this region, although
their application in the case of the JLC/NLC (for which the
inter-train repetition rate is 1.4 nsec) may require the development
of fast readout. If backgrounds are particularly high, particularly
if detectors are to be mounted below 25 mrad, it may be advantageous
to replace the silicon pad detectors with active pixel sensors. The
TESLA collaboration, which is considering calorimetry as close as
6 mrad from the beam axis, is also exploring the possibility of
using diamond instead of silicon. Finally, the Asian studies have
included work on an innovative
`3D Pixel' detector\cite{3D}, being developed at Stanford and the University of
Hawaii, which would detect electron-positron pairs from the beam-beam
interaction to within 10 mrad of the beam axis. This system would act
as a beam profile monitor, with sensitivity to various beam geometry
parameters.

The value of dedicated particle identification, somewhere in the
region inside the calorimeter, has not been fully explored. None of
the baseline designs includes such a device, but the case for its
omission remains to be clearly made. In fact, slow, heavy particles
are a fairly common signature of theories which extend the Standard
Model, suggesting that a time-of-flight system may be worthy of
consideration.

The identification of a technology for the muon system has yet to be
explored in a comprehensive way. Finally, the task of building
large solenoidal magnets with fields of up to 5T will certainly
be challenging, and require substantial development effort.

\section{Summary}

The complementarity of the LC program to that of the LHC relies
on a number of general factors: the definitive studies enabled
by precision tracking and vertexing and the a-priori
knowledge of the cms frame and energy, the accurate
reconstruction of hadronic final states, and the openness of the
LC trigger. All of these would benefit from systems which
perform substantially better than their existing counterparts
at LEP and the SLC, motivating a vigorous
program of research and development.
This program needs to include the development and exploitation of
sophisticated simulation tools, in order that the hardware development
effort be sufficiently informed. This research and development program --
both simulation studies and hardware development --
should become an important focus for the worldwide particle physics
community as the prospects for building a high-energy Linear
Collider improve.

%
%

%
%



\end{document}